\documentclass[mathleft
]{an}
\usepackage{graphicx}
\usepackage{times}
\def\e{$\pm$}  

\begin{document}

\Pagespan{189}{195}
\Yearpublication{2015}%
\Yearsubmission{2014}%
\Month{11}%
\Volume{336}%
\Issue{2}%
\DOI{10.1002/asna.201412142}  %

\title{Difference between the optical flickering colours of cataclysmic variables and symbiotic recurrent novae \,\thanks{Based on
    observations obtained  at National Astronomical Observatory Rozhen and Belogradchik Observatory, Bulgaria}}

\author{R. Zamanov\inst{1}\fnmsep\thanks{Corresponding author:
  \email{rkz@astro.bas.bg}\newline}
\and  S. Boeva\inst{1}
\and  G. Latev\inst{1}
\and  K. A. Stoyanov\inst{1}  
\and  S. V. Tsvetkova\inst{1}
}
\titlerunning{Flickering of cataclysmic variables and recurrent novae}
\authorrunning{Zamanov, Boeva, Latev et al.}
\institute{
Institute of Astronomy and National Astronomical Observatory, 
Bulgarian Academy of Sciences,  72 Tsarigradsko shose, BG-1784, Sofia, Bulgaria 
}

\received{2 September 2014}
\accepted{2014 Dec 04}
\publonline{2015 Mar 01}


\abstract{%
    We performed simultaneous observations in 3 bands $(UBV)$ of the flickering variability 
     of  the recurrent novae  RS Oph and T~CrB at quiescence. 
     Using new and published data,  
     we compare the colours of the flickering in cataclysmic variables  and  symbiotic recurrent novae.
     We find a difference between 
     the colours of the flickering source in these two types of accreting white dwarfs.
     The detected difference is highly significant with  $ p-value \approx 2 \times 10^{-6}$  for the distributions of 
     $(U-B)_0$ colour 
     and $p \approx 3 \times 10^{-5}$ on (U-B) versus (B-V) diagram.  \\
     The possible physical reasons are briefly discussed. 
     The data are available upon request from the authors.  }
\maketitle

\section{Introduction}
Cataclysmic variables (CVs) are semi-detached interacting binary systems containing an accreting white
dwarf  primary and a mass-losing, late-type, near or on the main sequence secondary star that fills its Roche lobe.
The accretion process can be either directly onto a strongly magnetic white dwarf or by
way of an intervening accretion disc.
The orbital periods for such  systems are most commonly between $\sim$ 80
minutes and several hours (e.g.  Warner 2003).

The {\it Recurrent Novae} (RNe) are previously recognized classical novae that repeat their outbursts.
RNe are ordinary novae systems for which 
the recurrence time scale happens to be  from a decade to a century. 
RNe are binary stars where matter accretes
from a donor star onto the surface of a white dwarf, where
the accumulated material will eventually start a thermonuclear explosion that makes the nova eruption (e.g. Anupama 2008; Schaefer 2010). 
The two RNe observed here (T~CrB and RS~Oph) belong to the group of the RNe with red giant companions 
and with very long periods comparable to one year, $P_{orb} = 227.6$~d for T~CrB (Fekel et al. 2000)
and  $P_{orb} =  453.6$~d for RS~Oph (Brandi et al. 2009). 
This type of nova is also referred to as a symbiotic recurrent nova, SyRN (e.g. Bode 2010, Shore et al. 2011). 
T~CrB and RS~Oph are also classified as symbiotic stars, because the mass donor is 
a red giant. 

The flickering (stochastic light variations on timescales of
a few minutes with amplitude of a few$\times0.1$ magnitudes)
is a variability observed in the three main types of binaries that contain white dwarfs 
accreting  material from a companion mass-donor star:  
cataclysmic variables (CVs), supersoft X-ray binaries,  
and symbiotic stars (Sokoloski 2003). Flickering is also visible in X-ray binaries and Active Galactic Nuclei at X-ray
wavelengths (e.g. Uttley \& McHardy 2001). Interestingly, the cataclysmic variable MV~Lyr displays rms-flux relation, which is similar to that 
from accreting black  holes (Scaringi et al. 2012).

The location of the flickering light source is supposed to be 
(1) at the outer edge of the accretion disk, where  the gas stream from the secondary hits the disk (Warner \& Nather 1971);
(2) at the inner edge of the disc and the boundary layer between the disk and  the central white  dwarf (Bruch 2000);
(3) inside the accretion disc itself (Baptista \& Bortoletto 2004).

The first clues that the colours of the flickering depend on the type of the source
can be found in  Fig.6 of Zamanov et al. (2010a). 
Here we report new UBV observations of the flickering variability of 
RS~Oph and T~CrB  and find  statistically significant difference 
between the colour of the flickering of  CVs and SyRNe.

\section{Observational data}

\subsection{New observations}
\label{obs.new}

Our new observations are obtained by the 2.0~m RCC, the 60~cm and the 50/70~cm Schmidt 
telescopes of the Bulgarian National Astronomical Observatory  Rozhen, located in the Rhodope
mountain range.  We also made use of the 60~cm telescope
of the  Belogradchik Astronomical Observatory, located in the vicinity of the Belogradchik Rocks. 
All the telescopes are equipped with CCD cameras. 
The 2.0 m RCC telescope possesses a dual-channel focal reducer (Jockers et al.  2000) and 
observes simultaneously in two bands - $U$ (blue channel) and $V$ (red channel). 
The journal of observations is given in Table~\ref{Tab.obs}.
The data are taken simultaneously using two or three telescopes. 
The only exception  is the run 20120613, when  the 60 cm Rozhen telescope observed 
in repeating U, B and V bands.

\begin{figure}
 \vspace{8.0cm}  
  \includegraphics{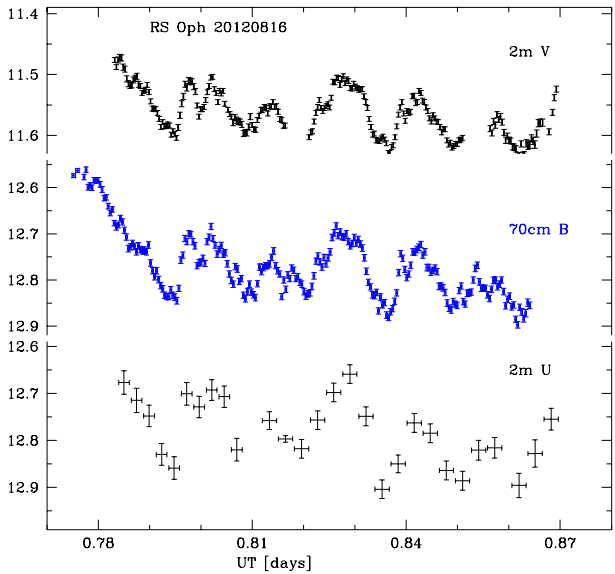}      
  \caption[]{Flickering variability of RS~Oph in the $UBV$ bands  on 2012 August 16.
  }		    
\label{fig.examp}     
\end{figure}

All the CCD images have been bias subtracted and flat fielded, and standard 
aperture photometry has been performed. The data reduction and aperture photometry 
are done with {\sc iraf} and have been checked with alternative software packages. 
For RS~Oph and T~CrB, the comparison stars of Henden \& Munari (2006) have been used.  
An example of our observations is given in Fig.\ref{fig.examp}. 


\subsection{Data from literature}
\label{obs.old}

Bruch (1992) measured the  colours of the flickering light source
of 12 CVs  (RX~And, AE~Aqr, V603~Aql, BV~Cen, WW~Cet, SS~Cyg, HR~Del, AH~Her, EX~Hya, WX~Hyi, V426~Oph, GK~Per)
and two RNe (RS~Oph and T~CrB).  
His data included 36 measurements of $(B-V)_0$   and  20 of $(U-B)_0$ colour.

Additionally we also made use of our own
partly published observations of 
four CVs:  
V425~Cas (Tsvetkova et al. 2010),
MV~Lyr (Boeva et al. 2011),
V794~Aql (Latev et al. 2011),
KR~Aur (Boeva et al. 2012)
and of two SyRNe: RS~Oph  (Zamanov et al. 2010a)  and T~CrB (Zamanov et al. 2010b),

\section{Differences between SyRNe and CVs}
\label{Diff}


\subsection{Colours of the flickering source} 

To calculate the colours of the flickering light source 
we follow Bruch (1992).
In this method  is considered  that  the flickering source is 100\% modulated and that all the variability 
during each night is due to flickering.  Following the  Bruch's (1992) method 
we calculate the flux of the flickering light source
for each band, using our UBV observations  and  Bessel (1979) calibration 
for the fluxes of a zero-magnitude star. 
After that we  transform the fluxes  into magnitudes,
correct them for the interstellar extinction, and calculate $(U-B)_0$ and $(B-V)_0$ colours of the flickering source. 
The flickering colours are computed on a timescale of $\approx \! 1$ hour.
Because usually the different bands 
do not start or end at the same time, 
we cut them (before the calculations) to the same time interval.
This is done for the new data and for our published observations as well.

We use the  reddening law from Fitzpatrick (1999).  
We adopt reddening 
$E(B-V) = 0.0 \pm 0.05$ for MV~Lyr, $E(B-V) = 0.05  \pm 0.05$ for KR~Aur (Verbunt 1987),  
$E(B-V) = 0.2$ for V794~Aql (Godon et al. 2007),
$E(B-V) =  0.73 \pm 0.06 $ for RS~Oph   (Snijders 1987), and  
$E(B-V) = 0.14 \pm 0.05$ for T~CrB (Parimucha \& Va{\v n}ko 2006).

The results are presented in Table~\ref{Tab1}, where  
the first column gives  the date of observation in format YYYYMMDD, 
the second --  minimum and maximum B band magnitude during the observation,   
the third and the fourth -- the $(B-V)_0$ and $(U-B)_0$ colours of the flickering sources, corrected 
for the interstellar extinction with the corresponding errors.
The  errors are evaluated from the accuracy of the photometry. 
They  depend  on the brightness of the object, and also on the flickering amplitude
(e.g. if the amplitude of the flickering is large, we obtain good flickering colours even if the object is 
with lower brightness). 



\subsection{Difference in  $(U-B)_0$  colours}
\label{UB}

\begin{figure}  
  \vspace{7.0cm}  
  \includegraphics{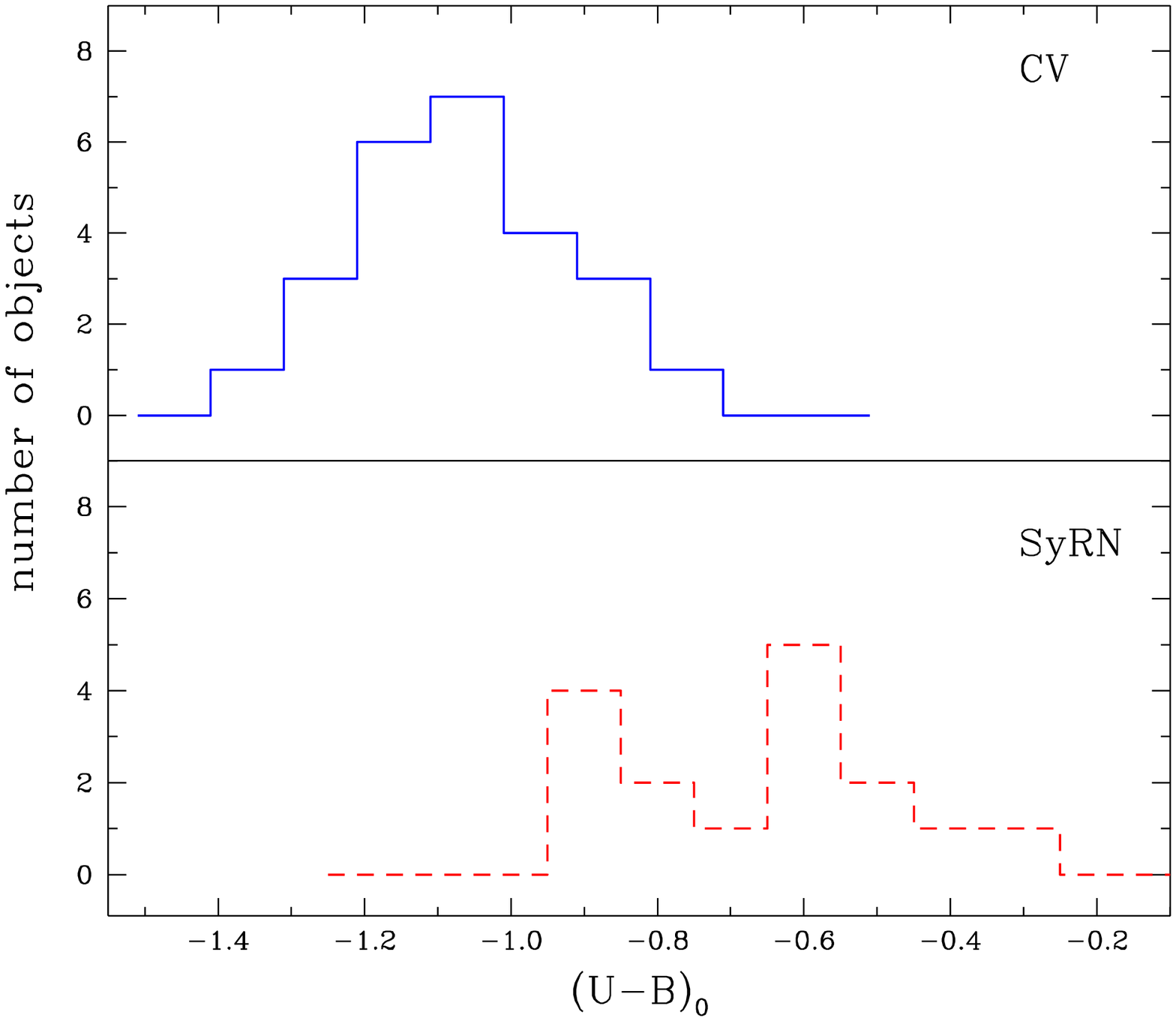}   
  \caption[]{Histograms showing the distribution of
  $(U-B)_0$  colour of the flickering source of CVs (blue solid line) and SyRNe (red dashed).   }		    
  \label{figUB}  
  \vspace{7.0cm}  
  \includegraphics{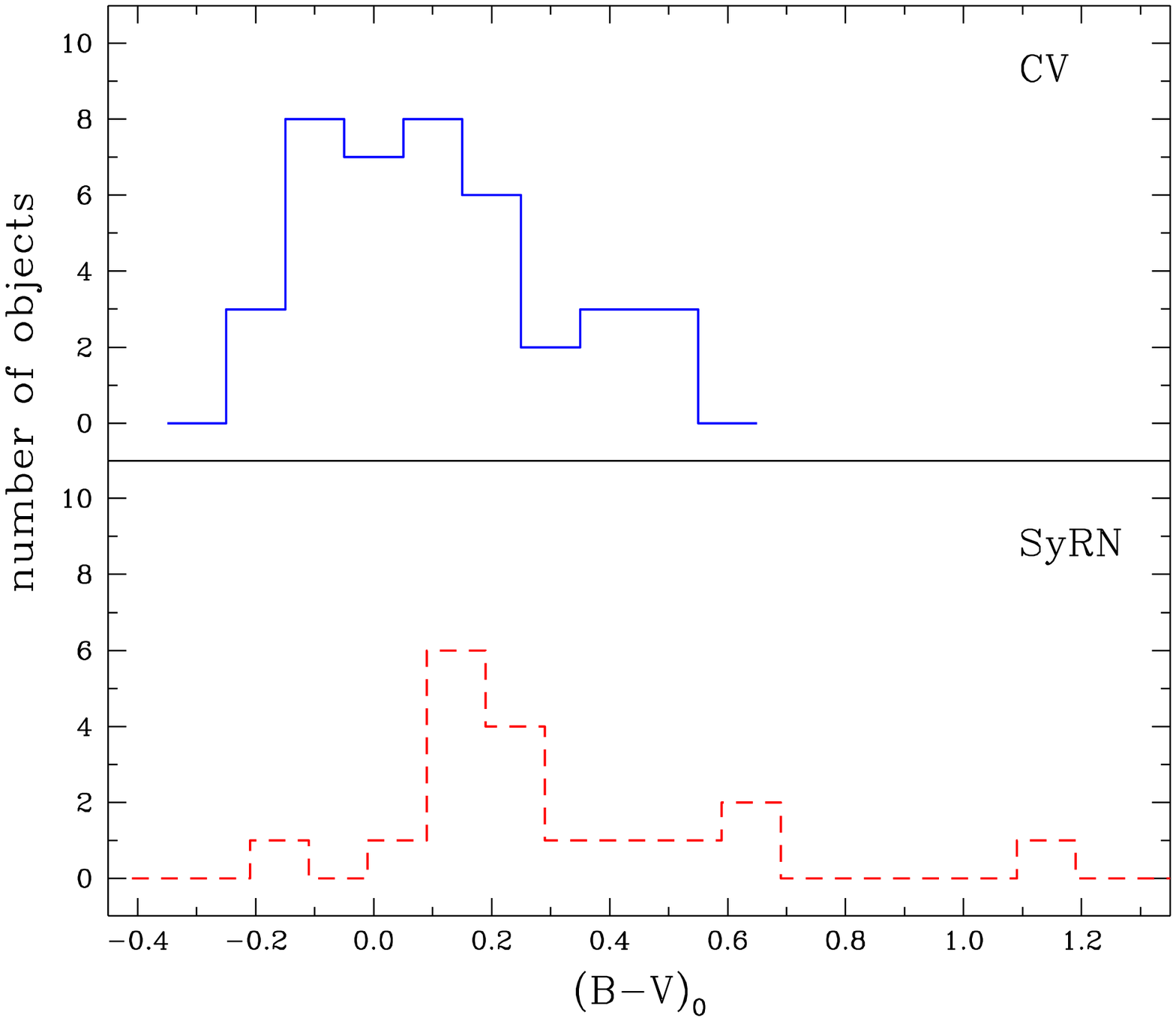}   
  \caption[]{Histograms showing the distribution of
  $(B-V)_0$  colour of the flickering source  of CVs (blue solid line) and SyRNe (red dashed).  }		    
  \label{figBV}        
\end{figure}

Fig.~\ref{figUB}  shows the histograms for  $(U-B)_0$
colours of the flickering light sources of CVs (solid line) and SyRNe (dashed line).
The Kolmogorov-Smirnov test compares the cumulative distributions of two data sets
($N_1=26$, $N_2=16$).
The test gives maximum difference between the cumulative distributions  $D=0.80$
and a probability of $p = 2.1  \times 10 ^{-6}$ that the  two data
sets are drawn from the same parent population.
 
The  Wilcoxon-Mann-Whitney U-Test checks the hypothesis 
whether two sample populations have the same median of distribution. 
It gives  $p= 4.8  \times 10 ^{-7}$. This value
is  better than  that  produced by the Kolmogorov-Smirnov test and confirms  the result.
The values of $p << 10^{-3}$ indicate that there is highly significant difference in  $(U-B)_0$ colour of the flickering 
between SyRNe and CVs.

\begin{figure}  
  \vspace{8.2cm}  
  \includegraphics{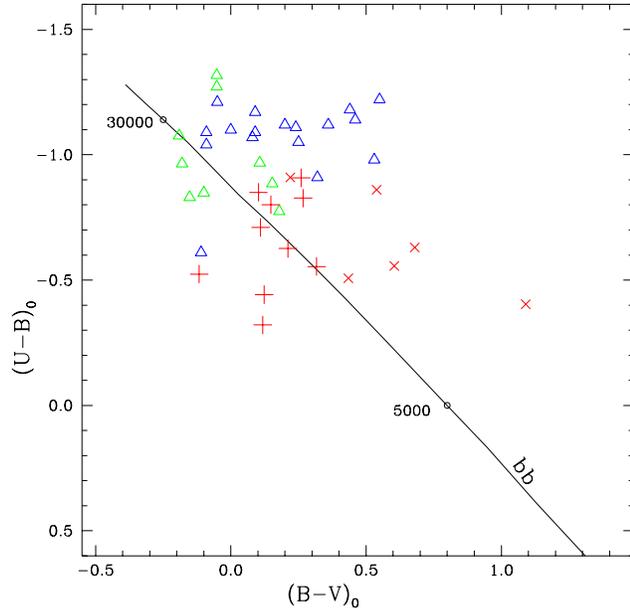}   
  \caption[]{ Positions of the flickering light 
  source on a $(U-B)$ vs $(B-V)$ diagram. The solid line represents the blackbody.
  RS~Oph  is marked with  (red) plus symbols,   T~CrB -- with (red) crosses.
  The blue (Bruch 1992) and green (our data from Table~\ref{Tab1}) triangles are the CVs.
  }		    
  \label{2C}     
\end{figure}	    

\subsection{$(B-V)_0$  colour}

Fig.~\ref{2C} shows the histograms for $(B-V)_0$  colours of the flickering light sources 
of CVs (solid line) and SyRNe (dashed line).
The Kolmogorov-Smirnov test on  the histograms in Fig.~\ref{figBV}  ($N_1=40$ and $N_2=18$) gives
a probability  $p=0.006$  (Kolmogorov-Smirnov statistic $D=0.46$). 
The  Wilcoxon-Mann-Whitney U-Test for $(B-V)_0$ 
gives $p= 3 \times 10 ^{-3}$.

Both tests  point to a statistically significant difference in  the $(B-V)_0$ colours 
at a level of 0.01.  This value is not as high as that 
in Sect.~\ref{UB} and indicates that the difference between  CVs and SyRN is considerably more pronounced  
in the $(U-B)_0$ colour, than in $(B-V)_0$.

\subsection{Two-colour diagram}

Fig.~\ref{2C} represents a two-colour diagram, 
$(U-B)_0$ versus $(B-V)_0$ for the flickering light source in a number of CVs  and SyRNe.
In this figure, the cataclysmic variables are given with
triangles: 17 blue triangles denote the data taken  from  Bruch (1992) 
and 9 green triangles represent our measurements (Table~\ref{Tab1}).
RS~Oph is represented by 10 red plus symbols -- 1 data point from Bruch (1992) and 
9 are our measurements.  T~CrB is represented by 6 red crosses, 2 of them are from Bruch (1992) and 
4 are our observations. The solid line represents the blackbody radiation.

The two samples, CVs and SyRNe, have $N_1=26$ and $N_2=16$ data points, respectively.
Even by eye it is visible that the two distributions occupy different regions on this 
diagram. 


Using the data  in Fig.\ref{2C}, we compare colours of the 
flickering of the SyRNe with those of the CVs, by
a two-dimensional Kolmogorov-Smirnov test (Peacock  1983; Fasano \& Franceschini 1987).
The test gives a probability of $3 \times 10^{-5}$ that both distributions are
extracted from the same parent population. This is a highly significant value, which indicates that in this diagram 
there are  statistically significant differences between the flickering of these two 
classes of accreting white dwarfs.

\section{Discussion}

The results of the statistical tests in  Sect.\ref{Diff} confirm with higher level of significance
($p \sim 10^{-6}$) our early findings ($p \sim 2 \cdot 10^{-3}$) based on a 
smaller amount of data (Zamanov et al. 2010a). 
We performed experiments to use different values of  $E(B-V)$
and the interstellar extinction law  (e.g. Cardelli, Clayton \& Mathis 1989, and other possibilities 
included in  NASA/IPAC extinction calculators). They change the $p-values$ by less than  50\%. 
We conclude that the two SyRNe have indeed different colours of the flickering 
in comparison with the CVs.

\subsection{Parameters of the binary systems}

There are a few important differences between the classical CVs and the SyRNe, which 
could be the reason for the detected differences (Sect.~{\ref{Diff}) in the colours of the flickering sources:

1. the orbital periods of CVs are less than one day, while for the SyRNe 
they are longer than 100~d.

2. in CVs  mass accretion rates are $\dot M_{acc} \sim$ $1 \times 10^{-10} - 7 \times 10^{-9}$~M$_\odot$~yr$^{-1}$ (Echevarr{\'{\i}}a 1994). 
The SyRNe accrete at  higher rates: 
T~CrB at $\dot M_{acc} \approx 2.3 \times 10^{-8}$~M$_\odot$~yr$^{-1}$  (Selvelli, Cassatella \& Gilmozzi 1992),
RS~Oph  at $\dot M_{acc} \approx 2 \times 10^{-8}$~M$_\odot$~yr$^{-1}$  (Nelson et al. 2011).

3. in the CVs, the average value of the  white dwarf mass is $M_{WD} = 0.83 \pm 0.23$ M$_\odot$, and generally  $M_{WD} \le 1.1 $~M$_\odot$
(Zorotovic,  Schreiber, \&  G\"ansicke  2011;  Savoury et al. 2011).
The masses of the white dwarfs in SyRNe are close to the Chandrasekhar limit  
$M_{WD} = 1.37 \pm 0.13$~M$_\odot$  in T~CrB (Stanishev et al.  2004, Mukai et al. 2013),  
$M_{WD} =  1.2 -1.4 $~M$_\odot$  in RS~Oph (Brandi et al.  2009). 

4. As a consequence of 3., the white dwarf radius in RNe is expected to be smaller. 
Using Eggleton's  mass-radius relation as quoted by Verbunt \& Rappaport (1988), we estimate that in 
T~CrB and RS~Oph, 
the white dwarf radius should be smaller by a factor of 2. 
This could play an important role if the flickering is connected with the boundary layer between the accretion disk and 
the white dwarf
(see Bruch \& Duschl 1993). 

It is worth noting that 
  (i) the mass donors in SyRNe are red giants rather than main-sequence stars in CVs. 
Although the temperatures of both are comparable, the contribution of the red giants 
to the  quiet light in the optical bands could  be larger. 
However this will not affect our results, because the method applied 
here uses only the variable part of the light curve; 
(ii) the optical flux in most symbiotic stars is
affected by light from the red giant, which does not change on
short time-scales (see also  Sect 6.2 in Sokoloski, Bildsten \& Ho 2001).




\subsection{Models of the flickering}
\label{mod}

Yonehara, Mineshige \& Welsh (1997) proposed a model in which light fluctuations are produced by occasional 
flare-like events and subsequent  avalanche flow in the accretion disk atmospheres. 
Ribeiro \& Diaz (2006) simulated the  flickering  as a set of discrete flares on the accretion disk. Each flare 
is generated at a random position inside a pre-defined region. 
In these models the detected  
difference in the colours of the flickering   should  be associated with the temperature
and distribution of the flare-like events.

Dobrotka et al. (2010) proposed that the aperiodic variability is produced by turbulent elements in the disc.
In this model the detected distinction in Fig.\ref{2C}, could be due to a difference 
in the temperatures of the material in the turbulent elements 
as a result of the higher mass transfer rate and larger accretion disks  in the SyRNe. 

Scaringi (2014) developed a physical model for the flickering variability in CVs.
Many statistical properties of the flickering are explained with the fluctuating accretion 
disc model. 
In this model the detected difference  could be connected to the manner in which 
the disk is  fluctuating  at higher and lower mass accretion rates. 

In the CVs, the ultraviolet line diagnostics demonstrates that the accretion disk 
losses mass in the form of accretion disk winds (e.g. Vitello \& Shlosman 1993),
while in T~CrB and RS~Oph the opposite process (accretion from stellar wind) plays an important role in 
the accretion.  The stellar wind capture contributes about 20\%  to the total  mass transfer rate in T~CrB 
(Selvelli et al. 1992), and is also important in  RS~Oph (Wynn 2008).  
The presence/non-presence of disk wind could change the properties of flares and fluctuations. 

\subsection{Place of the flickering}

In CVs the average $(U-B)_0$ colour of the flickering  corresponds to a blackbody temperature $\sim 20000$~K, 
and the average $(B-V)_0$ to $\sim 10000$~K. In SyRNe both colours give a  temperature $\sim 9000$~K. 

The radial temperature profile of a steady-state accretion disk is:
\begin{equation}
T_{eff}^4=\frac{3 G \dot M_{acc} M_{WD}}{8 \pi \sigma R^3} 
\left[ 1-\left(\frac{R_{WD}}{R}\right)^{1/2} \right] ,
\end{equation}
where $G$ is the gravitational constant, $\sigma$ is the Stefan-Boltzmann constant, 
$R_{WD}$ is the white dwarf radius,
$R$ is the radial distance from the white dwarf.
Using the parameters for SyRNe, a temperature  $\sim$9000 K 
should be achieved at a distance $R \! \sim \! 0.5$ $R_\odot$ from the white dwarf, which corresponds to $\sim \! 120$~$R_{WD}$.
For the CVs the colours give a location of the flickering 
at $R \sim 0.08 - 0.20$ $R_\odot$ from the white dwarf, 
which corresponds to $8-20$~$R_{WD}$.
Perhaps in CVs, as a result of lower mass accretion rate 
and  lower white dwarf mass, the flares and/or fluctuations (see Sect.\ref{mod})
are ignited closer to the white dwarf in comparison with SyRNe, which results in different colours of the flickering. 

\section{Conclusions}
We find a statistically highly-significant difference between the colours of the flickering source in cataclysmic variables 
and symbiotic recurrent novae. 
This difference is  likely to be connected with the mass accretion rate and white dwarf mass
and should be useful to test the theoretical models of flickering. 

\acknowledgements    
Part of this work was supported by the OP "HRD", ESF and Bulgarian  Ministry of Education and Science
    (BG051PO001-3.3.06-0047).  
The authors thank the referee, Simone Scaringi, for his constructive comments.




\begin{table*}
\caption{CCD observations of flickering. The following are given in the table:  date,  UT start and UT end of the run, the telescope, band, 
exposure time, number of CCD images, average magnitude in the corresponding band, minimum-­maximum magnitudes in each band, standard
deviation of the mean and typical observational error. }
\begin{center}
\begin{tabular}{lclcccccllccll}
\\
\hline
date           &     UT       & telescope& band & exp-time & N$_{pts}$  & Average &  Min.-- Max.     & St. dev.  & Error & \\
               &  start-end   &          &      &  (s)     &            &  (mag)  &  (mag)-(mag)     & (mag)     & (mag) & \\
\hline
\\

{\bf RS~Oph} \\                   
2008 Jul 06$^a$ & 19:46-21:28  & 2.0 m Roz      & U & 300        & 16	  & 12.546  & 12.450 - 12.721  & 0.074     & 0.011 & \\
	        &	       & 50/70 cm Sch   & B & 20,60,100  & 48	  & 12.471  & 12.380 - 12.615  & 0.052     & 0.005 & \\
	        &	       & 2.0 m Roz      & V & 30         & 160	  & 11.232  & 11.160 - 11.333  & 0.037     & 0.007 & \\
2009 Jul 21$^a$ & 20:57-21:33  & 50/70 cm Sch   & U & 120        & 16	  & 12.016  & 11.863 - 12.166  & 0.098     & 0.013 & \\
	        &	       & 60 cm Roz      & B & 40         & 29	  & 12.114  & 11.991 - 12.277  & 0.070     & 0.007 & \\
	        &	       & 60 cm Bel      & V & 30         & 39	  & 11.069  & 10.895 - 11.236  & 0.081     & 0.004 & \\
2009 Jul 23$^a$ & 20:52-21:30  & 2.0 m Roz      & U & 120        & 10	  & 11.719  & 11.606 - 11.845  & 0.082     & 0.028 & \\
	        &              & 50/70 cm Sch   & B & 30         & 38	  & 11.979  & 11.857 - 12.099  & 0.067     & 0.007 & \\
	        &	       & 2.0 m Roz      & V & 10         & 100	  & 10.978  & 10.863 - 11.082  & 0.064     & 0.008 & \\
2012 Apr 27$^b$ & 00:28-01:34  & 50/70 cm Sch   & U & 60,120     & 35	  & 12.050  & 11.850 - 12.244  & 0.091     & 0.011 & \\
	        &	       & 60 cm Roz      & B & 60         & 53	  & 12.112  & 11.934 - 12.271  & 0.074     & 0.006 & \\
	        &	       & 60 cm Bel      & V & 20         & 97	  & 10.992  & 10.882 - 11.120  & 0.053     & 0.006 & \\
2012 Jun 13$^b$ & 21:45-23:38  & 60 cm Roz      & U & 120        & 34	  & 12.599  & 12.407 - 12.762  & 0.089     & 0.014 & \\
	        &	       & 60 cm Roz      & B & 20         & 34	  & 12.531  & 12.380 - 12.671  & 0.086     & 0.012 & \\
	        &	       & 60 cm Roz      & V & 10         & 34	  & 11.376  & 11.274 - 11.492  & 0.063     & 0.007 & \\
2012 Aug 15$^b$ & 18:44-20:30  & 2.0 m Roz      & U & 90         & 55	  & 13.089  & 12.659 - 13.538  & 0.303     & 0.019 & \\
	        &	       & 60 cm Roz      & B & 60         & 65	  & 12.949  & 12.624 - 13.208  & 0.211     & 0.009 & \\
	        &	       & 2.0 m Roz      & V & 10         & 53	  & 11.704  & 11.453 - 11.908  & 0.157     & 0.008 & \\
2012 Aug 16$^b$ & 18:48-20:48  & 2.0 m Roz      & U & 180,240    & 26	  & 12.783  & 12.659 - 12.904  & 0.071     & 0.022 & \\
	        &	       & 50/70 cm Sch   & B & 30         & 217	  & 12.775  & 12.584 - 12.897  & 0.064     & 0.007 & \\
	        &	       & 2.0 m Roz      & V & 10         & 217	  & 11.563  & 11.472 - 11.633  & 0.038     & 0.008 & \\
2013 Aug 12$^b$ & 19:09-21:30  & 2.0 m Roz      & U & 180        & 34	  & 11.996  & 11.766 - 12.171  & 0.121     & 0.009 & \\
	        &	       & 50/70 cm Sch   & B & 20         & 412	  & 12.121  & 11.916 - 12.270  & 0.085     & 0.012 & \\
	        &	       & 2.0 m Roz      & V & 3,5        & 335	  & 11.060  & 10.897 - 11.203  & 0.070     & 0.005 & \\
2013 Aug 13$^b$ & 20:40-21:22  & 2.0 m Roz      & U & 180        & 11	  & 12.580  & 12.522 - 12.645  & 0.044     & 0.011 & \\
                &              & 50/70 cm Sch   & B & 15,20      & 84     & 12.579  & 12.476 - 12.659  & 0.042     & 0.010 & \\
                &              & 2.0 m Roz      & V & 3	         & 118	  & 11.441  & 11.378 - 11.503  & 0.028     & 0.004 & \\
\\
{\bf T~CrB} \\
2009 Jan 20$^a$ & 02:27-04:07  & 2.0 m Roz      & U & 180        & 28	  & 11.747  & 11.517 - 11.921  & 0.118     & 0.025 & \\
	        &              & 50/70 cm Sch   & B & 10         & 240	  & 11.199  & 11.108 - 11.269  & 0.035     & 0.011 & \\
	        &              & 2.0 m Roz      & V & 5	         & 559	  & 9.916   & 9.878 - 9.952    & 0.015     & 0.009 & \\
2009 Feb 27$^a$ & 00:15-01:40  & 2.0 m Roz      & U & 180        & 24	  & 12.330  & 12.162 - 12.465  & 0.089     & 0.028 & \\
	        &              & 50/70 cm Sch   & B & 40         & 24	  & 11.482  & 11.438 - 11.521  & 0.021     & 0.005 & \\
	        &              & 2.0 m Roz      & V & 7	         & 490	  & 10.168  & 10.138 - 10.191  & 0.009     & 0.007 & \\
2010 Apr 30$^b$ & 18:46-20:25  & 60 cm Roz      & U & 120        & 27	  & 12.003  & 11.765 - 12.196  & 0.115     & 0.006 & \\
	        &              & 60 cm Roz      & B & 30         & 30	  & 11.405  & 11.339 - 11.457  & 0.027     & 0.007 & \\
2010 Apr 30$^b$ & 20:30-22:10  & 60 cm Roz      & B & 30         & 110	  & 11.410  & 11.350 - 11.469  & 0.027     & 0.012 & \\
	        &              & 60 cm Roz      & V & 10         & 109	  & 10.042  & 9.992 - 10.073   & 0.015     & 0.008 & \\
2011 Feb 11$^b$ & 01:04-03:44  & 50/70 cm Sch   & U & 120        & 86	  & 12.790  & 12.686 - 12.857  & 0.037     & 0.006 & \\
                &	       & 60 cm Roz      & B &  40	 & 121	  & 11.801  & 11.781 - 11.817  & 0.007     & 0.004 & \\
                &              & 60 cm Roz      & V &  20        & 121    & 10.343  & 10.330 - 10.356  & 0.006     & 0.003 & \\
\\
{\bf KR~Aur} \\
 2009 Jan 20$^a$ & 19:06-22:21 & 2.0 m Roz      & U & 300        &  25	  & 17.227  & 16.310 - 17.730  & 0.417    & 0.050 & \\
	         &  	       & 50/70 cm Sch   & B & 300        &  28	  & 18.342  & 17.510 - 18.780  & 0.366    & 0.035 & \\
	         &  	       & 2.0 m Roz      & V & 300        &  27	  & 18.342  & 17.510 - 18.780  & 0.366    & 0.030 & \\ 
 2009 Feb 26$^a$ & 17:36-23:34 & 2.0 m Roz      & U & 300        &  49	  & 17.706  & 17.110 - 17.930  & 0.196    & 0.050 & \\ 
	         &   	       & 50/70 cm Sch   & B & 300        &  20	  & 18.720  & 17.210 - 18.890  & 0.073    & 0.025 & \\
	         &  	       & 2.0 m Roz      & V & 300        &  53	  & 18.720  & 17.210 - 18.890  & 0.073    & 0.020 & \\ 
 2010 Dec 31$^b$ & 17:59-20:03 & 2.0 m Roz      & U & 300        &  22	  & 13.241  & 13.080 - 13.410  & 0.124    & 0.030 & \\
	         &  	       & 50/70 cm Sch   & B & 60	 &  117	  & 14.262  & 14.032 - 14.487  & 0.124    & 0.020 & \\
                 &             & 2.0 m Roz      & V & 30         &  237   & 14.082  & 13.853 - 14.319  & 0.111    & 0.010 & \\ 
\\
\hline
\end{tabular}															     
\end{center}															     
\label{Tab.obs}															     
\end{table*}							
\addtocounter{table}{-1}
\begin{table*} 
\caption{Continuation.}
\begin{center}
\begin{tabular}{lllcccccllccll}
\\
\hline
date           &     UT       & telescope& band & exp-time & N$_{pts}$  & Average &  Min.- Max.      & St. dev. & Error & \\
               &  start-end   &          &      &  (s)     &            &  (mag)  &  (mag)-(mag)     & (mag)    & (mag) & \\
\hline
\\
{\bf MV~Lyr} \\
 2009 Jul 22$^a$ & 20:04-21:27 & 2.0 m Roz     & U & 120	    &  30	  & 12.796  & 12.580 - 12.916  & 0.089    & 0.031 & \\
	         &	       & 50/70 cm Sch  & B &  60	    &  73	  & 13.702  & 13.502 - 13.805  & 0.083    & 0.007 & \\
	         &	       & 2.0 m Roz     & V & 30	            &  144	  & 13.657  & 13.479 - 13.777  & 0.073    & 0.005 & \\ 
 2010 Aug 15$^a$ & 22:37-23:59 & 50/70 cm Sch  & U & 180	    &  27	  & 12.135  & 12.000 - 12.221  & 0.062    & 0.020 & \\
	         &	       & 60 cm Roz     & B & 120	    &  25	  & 12.946  & 12.810 - 13.029  & 0.054    & 0.030 & \\
	         &	       & 60 cm Bel     & V & 30,40,60       &  62	  & 12.912  & 12.767 - 12.985  & 0.050    & 0.005 & \\ 
\\
{\bf V794~Aql} \\
 2009 Jul 23$^a$ & 00:16-01:18 & 2.0 m Roz     & U & 240	     &  13	  & 15.264  & 15.129 - 15.458  & 0.089    & 0.060 & \\
	         &	       & 50/70 cm Sch  & B & 180	     &  20	  & 16.283  & 16.160 - 16.542  & 0.102    & 0.060 & \\
	         &	       & 2.0 m Roz     & V &  60	     &  53	  & 15.992  & 15.835 - 16.270  & 0.099    & 0.020 & \\ 
 2010 Aug 13$^a$ & 19:43-21:26 & 2.0 m Roz     & U & 300	     &  18	  & 14.580  & 14.450 - 14.723  & 0.092    & 0.016 & \\
	         &	       & 50/70 cm Sch  & B & 180	     &  32	  & 15.357  & 15.226 - 15.514  & 0.078    & 0.012 & \\
	         &	       & 2.0 m Roz     & V &  90	     &  83	  & 15.055  & 14.941 - 15.178  & 0.067    & 0.007 & \\ 
\\
{\bf V425~Cas} \\
 2009 Jan 20$^a$ & 17:01-18:37 & 2.0 m Roz     & U &  20             &  25	  & 14.823  & 14.680 - 14.980  & 0.070    & 0.013 & \\
	         &  	       & 50/70 cm Sch  & B &  20             &  44	  & 15.516  & 15.370 - 15.650  & 0.062    & 0.020 & \\
	         &  	       & 2.0 m Roz     & V &  20             &  55	  & 15.343  & 15.220 - 15.470  & 0.055    & 0.020 & \\ 
 2009 Jul 23$^a$ & 00:20-01:30 & 2.0 m Roz     & U & 180             &  18	  & 14.084  & 13.953 - 14.228  & 0.082    & 0.006 & \\
	         &  	       & 50/70 cm Sch  & B & 180             &  32	  & 14.934  & 14.781 - 15.094  & 0.086    & 0.010 & \\
	         &  	       & 2.0 m Roz     & V & 60	             &  62	  & 14.719  & 14.549 - 14.866  & 0.080    & 0.010 & \\ 
\\
\hline
\end{tabular}															     
\end{center}	
\footnotetext{a}{$^a$ partly published observations, see Sect.\ref{obs.old} for references. \\ }
\footnotetext{b}{$^b$ new data }														     
\label{Tab.cont}															     
\end{table*}		


\begin{table}
\caption{Colours of the flickering light source, corrected for interstellar extinction.
}
\begin{tabular}{lccccccrllll}
\\
\hline
 object  &     $(B-V)_0$        &  $(U-B)_0$                 \\
 date    &    			                 	     \\
\hline 
\\
{\bf RS~Oph} \\                                                     	   
20080706  &   $+$0.123\e0.06  & $-$0.442\e0.05   \\   
20090721  &   $+$0.317\e0.04  & $-$0.554\e0.05   \\   
20090723  &   $+$0.102\e0.06  & $-$0.850\e0.15   \\   
20120427  &   $+$0.148\e0.05  & $-$0.801\e0.04   \\   
20120613  &   $+$0.211\e0.06  & $-$0.626\e0.08   \\   
20120815  &   $+$0.260\e0.04  & $-$0.908\e0.04   \\   
20120816  &   $-$0.117\e0.07  & $-$0.524\e0.16   \\   
20130812  &   $+$0.267\e0.04  & $-$0.827\e0.05   \\   
20130813  &   $+$0.118\e0.15  & $-$0.321\e0.15   \\   
\\
{\bf T~CrB} \\
20090120  &   $+$0.434\e0.13  & $-$0.507\e0.05   \\	
20090227  &   $+$0.604\e0.20  & $-$0.556\e0.06   \\	
20100430  &   $+$0.539\e0.30  & $-$0.860\e0.15   \\	
20110211  &   $+$1.090\e0.35  & $-$0.644\e0.14   \\	
\\
{\bf KR~Aur} \\
20090120  &   $-$0.052\e0.07  & $-$1.272\e0.08  \\    
20090226  &   $-$0.052\e0.11  & $-$1.324\e0.18  \\    
20101231  &   $+$0.179\e0.04  & $-$0.774\e0.08  \\    
\\
{\bf MV~Lyr} \\
20090722  &   $+$0.107\e0.04  & $-$0.968\e0.08  \\    
20100815  &   $-$0.100\e0.07  & $-$0.848\e0.15  \\    
\\
{\bf V794~Aql}\\
20090724  &   $+$0.153\e0.04  & $-$0.885\e0.07  \\    
20100813  &   $-$0.152\e0.05  & $-$0.830\e0.09  \\    
\\
{\bf V425~Cas}\\
20090120  &   $-$0.191\e0.12  & $-$1.076\e0.14  \\    
20090723  &   $-$0.180\e0.17  & $-$0.965\e0.18  \\    
\\
\hline
\\
\end{tabular}						 							      
\label{Tab1}															     
\end{table}															     


\begin{thebibliography}{44}

\bibitem[Anupama(2008)]{2008ASPC..401...31A} Anupama, G.~C.\ 2008, RS Ophiuchi (2006) and the Recurrent Nova Phenomenon, ASP Conference Series 401, 31 

\bibitem[Baptista \& Bortoletto(2004)]{2004AJ....128..411B} Baptista, R., \& Bortoletto, A.\ 2004, \aj, 128, 411 

\bibitem[Bessell(1979)]{1979PASP...91..589B} Bessell, M.~S.\ 1979, \pasp,  91, 589 

\bibitem[Bode(2010)]{2010AN....331..160B} Bode, M.~F.\ 2010, Astronomische Nachrichten, 331, 160 

\bibitem[Boeva et al.(2011)]{2011BlgAJ..16...23B} Boeva, S., Bachev, R., Tsvetkova, S., et al.\ 2011, Bulgarian Astronomical Journal, 16, 23 

\bibitem[Boeva et al.(2012)]{2012PASRB..11..369B} Boeva, S., Bachev, R., Antov, A., Tsvetkova, S., \& Stoyanov, K.~A.\ 2012, Publications of the Astronomical Society ''Rudjer Boskovic'', 11, 369 

\bibitem[Brandi et al.(2009)]{2009A&A...497..815B} Brandi, E., Quiroga, C., Miko{\l}ajewska, J., Ferrer, O.~E., \& Garc{\'{\i}}a, L.~G.\ 2009, A\&A, 497, 815 

\bibitem[Bruch(1992)]{1992A&A...266..237B} Bruch, A.\ 1992, A\&A, 266, 237 

\bibitem[Bruch(2000)]{2000A&A...359..998B} Bruch, A.\ 2000, A\&A, 359, 998 

\bibitem[Bruch \& Duschl(1993)]{1993A&A...275..219B} Bruch, A., \& Duschl, W.~J.\ 1993, A\&A, 275, 219 

\bibitem[Cardelli et al.(1989)]{1989ApJ...345..245C} Cardelli, J.~A., Clayton, G.~C., \& Mathis, J.~S.\ 1989, \apj, 345, 245 

\bibitem[Dobrotka et al.(2010)]{2010MNRAS.402.2567D} Dobrotka, A., Hric, L., Casares, J., et al.\ 2010, \mnras, 402, 2567 

\bibitem[Echevarr{\'{\i}}a(1994)]{1994RMxAA..28..125E} Echevarr{\'{\i}}a, J.\ 1994, Rev. Mexicana Astron. Astrofis., 28, 125 

\bibitem[Fasano \& Franceschini(1987)]{1987MNRAS.225..155F} Fasano, G., \& Franceschini, A.\ 1987, \mnras, 225, 155 

\bibitem[Fekel et al.(2000)]{2000AJ....119.1375F} Fekel, F.~C., Joyce, R.~R., Hinkle, K.~H., \& Skrutskie, M.~F.\ 2000, \aj, 119, 1375 

\bibitem[Fitzpatrick(1999)]{1999PASP..111...63F} Fitzpatrick, E.~L.\ 1999, \pasp, 111, 63 

\bibitem[Godon et al.(2007)]{2007ApJ...656.1092G} Godon, P., Sion, E.~M., Barrett, P., \& Szkody, P.\ 2007, \apj, 656, 1092 

\bibitem[Henden \& Munari(2006)]{2006A&A...458..339H} Henden, A., \& Munari, U.\ 2006, A\&A, 458, 339 

\bibitem[Jockers et al.(2000)]{2000KFNTS...3...13J} Jockers, K., Credner, T., Bonev, T., et al.\ 2000, Kinematika i Fizika Nebesnykh Tel Supplement, 3, 13 

\bibitem[Latev et al.(2011)]{2011BlgAJ..17...80L} Latev, G., Boeva, S., Zamanov, R., Antov, A., Stoyanov, K., Petrov, B., Tsvetkova, S., Spassov, B. \ 2011, Bulgarian Astronomical Journal, 17, 80 

\bibitem[Mukai et al.(2013)]{2013IAUS..281..186M} Mukai, K., Sokoloski, J.~L., Nelson, T., \& Luna, G.~J.~M.\ 2013, IAU Symposium, 281, 186 

\bibitem[Nelson et al.(2011)]{2011ApJ...737....7N} Nelson, T., Mukai, K., Orio, M., Luna, G.~J.~M., \& Sokoloski, J.~L.\ 2011, \apj, 737, 7 

\bibitem[Parimucha \& Va{\v n}ko(2006)]{2006ASPC..349..309P} Parimucha, {\v S}., \& Va{\v n}ko, M.\ 2006, Astrophysics of Variable Stars, ASP Conference Series, Vol. 349, p. 309 

\bibitem[Peacock(1983)]{1983MNRAS.202..615P} Peacock, J.~A.\ 1983, \mnras, 202, 615 


\bibitem[Ribeiro \& Diaz(2006)]{2006Ap&SS.304..291R} Ribeiro, F.~M.~A., \& Diaz, M.\ 2006, \apss, 304, 291 

\bibitem[Savoury et al.(2011)]{2011MNRAS.415.2025S} Savoury, C.~D.~J., Littlefair, S.~P., Dhillon, V.~S., et al.\ 2011, \mnras, 415, 2025 

\bibitem[Scaringi et al.(2012)]{2012MNRAS.421.2854S} Scaringi, S., K{\"o}rding, E., Uttley, P., et al.\ 2012, \mnras, 421, 2854 

\bibitem[Scaringi(2014)]{2014MNRAS.438.1233S} Scaringi, S.\ 2014, \mnras, 438, 1233 

\bibitem[Schaefer(2010)]{2010ApJS..187..275S} Schaefer, B.~E.\ 2010, \apjs, 187, 275 


\bibitem[Selvelli et al.(1992)]{1992ApJ...393..289S} Selvelli, P.~L., Cassatella, A., \& Gilmozzi, R.\ 1992, \apj, 393, 289 

\bibitem[Shore et  al.(2011)]{2011A&A...527A..98S} Shore, S.~N., Wahlgren, G.~M., Augusteijn, T., et al.\ 2011, A\&A, 527, A98 

\bibitem[Snijders(1987)]{1987Ap&SS.130..243S} Snijders, M.~A.~J.\ 1987, \apss, 130, 243 

\bibitem[Sokoloski et al.(2001)]{2001MNRAS.326..553S} Sokoloski, J.~L., Bildsten, L., \& Ho, W.~C.~G.\ 2001, \mnras, 326, 553 

\bibitem[Sokoloski(2003)]{2003ASPC..303..202S} Sokoloski, J.~L.\ 2003, Symbiotic Stars Probing Stellar Evolution,  ASP Conference Series 303, 202 

\bibitem[Stanishev et al.(2004)]{2004A&A...415..609S} Stanishev, V., Zamanov, R., Tomov, N., \& Marziani, P.\ 2004, A\&A, 415, 609 

\bibitem[Tsvetkova et al.(2010)]{2010POBeo..90..183T} Tsvetkova, S., Boeva, S., Zamanov, R., et al.\ 2010, Publications de l'Observatoire Astronomique de Beograd, 90, 183 

\bibitem[Uttley \& McHardy(2001)]{2001MNRAS.323L..26U} Uttley, P., \& McHardy, I.~M.\ 2001, \mnras, 323, L26 

\bibitem[Verbunt(1987)]{1987A&AS...71..339V} Verbunt, F.\ 1987, A\&As, 71, 339 

\bibitem[Verbunt \& Rappaport(1988)]{1988ApJ...332..193V} Verbunt, F., \& Rappaport, S.\ 1988, \apj, 332, 193 

\bibitem[Vitello \& Shlosman(1993)]{1993ApJ...410..815V} Vitello, P., \& Shlosman, I.\ 1993, \apj, 410, 815 

\bibitem[Warner \& Nather(1971)]{1971MNRAS.152..219W} Warner, B., \& Nather, R.~E.\ 1971, \mnras, 152, 219 

\bibitem[Warner(2003)]{2003cvs..book.....W} Warner, B.\ 2003, Cataclysmic Variable Stars, Cambridge University Press, Cambridge, UK  

\bibitem[Wynn(2008)]{2008ASPC..401...73W} Wynn, G.\ 2008, RS Ophiuchi (2006) and the Recurrent Nova Phenomenon, ASP Conference Series 401, p73

\bibitem[Yonehara et al.(1997)]{1997ApJ...486..388Y} Yonehara, A., Mineshige, S., \& Welsh, W.~F.\ 1997, \apj, 486, 388 

\bibitem[Zamanov et al.(2010)]{2010MNRAS.404..381Z} Zamanov, R.~K., Boeva, S., Bachev, R., et al.\ 2010a, \mnras, 404, 381 

\bibitem[Zamanov et al.(2010)]{2010ATel.2586....1Z} Zamanov, R., Boeva, S., Tsvetkova, S., \& Stoyanov, K.\ 2010b, The Astronomer's Telegram, 2586, 1 

\bibitem[Zorotovic et al.(2011)]{2011A&A...536A..42Z} Zorotovic, M., Schreiber, M.~R., G\"ansicke, B.~T.\ 2011, A\&A, 536, A42 

\end{thebibliography}
\end{document}